# Unveiling Mechanical Motions in Non-linear Optical Organic Micro Ring Resonators

Melchi Chosenyah,[1] Swaraj Rajaram,[1] Vladimir Novikov,[2] Anton Maydykovskiy[2], Ankur Khapre,[1] Tatiana Murzina,[2] Rajadurai Chandrasekar[1,✉] (ORCID: 0000-0003-3527-3984)

[1]Advanced Photonic Materials and Technology Laboratory, School of Chemistry and Centre for Nanotechnology, University of Hyderabad, Gachibowli, Hyderabad 500046, Telangana, India.
E-mail: r.chandrasekar@uohyd.ac.in

[2]Quantum Electronics Division, Department of Physics, Lomonosov Moscow State University, Leninskie Gory 1, Moscow 119991, Russia.

## Abstract

This work demonstrates the multifaceted dynamic motion and three-dimensional (3D) spatial manipulation of organic micro ring resonators (MRRs). The MRRs are fabricated via surface-tension-assisted self-assembly of 2,2'-((1E,1'E)-hydrazine-1,2-diylidenebis(methaneylylidene))bis(3,5-dibromophenol) (HDBP), exhibiting nonlinear optical (NLO) emission and frequency comb-type whispering gallery modes. Interestingly, the MRRs are micromechanically reconfigurable into various strained architectures, including lifting, transferring, vertical standing, axial spinning, and wheel-like rolling, using an atomic force microscopy cantilever tip. The MRRs retained their photonic traits throughout these dynamic motions, underscoring their mechanical robustness. Notably, the demonstration of axial spinning and rolling locomotion extends the manipulation capabilities beyond two-dimensional control, enabling complete 3D spatial control. These results establish a foundational platform for next-generation NLO mechanophotonic systems, where reconfigurable smart and soft photonic elements can be dynamically controlled with high spatial precision.

**Main**

Ring-shaped architectures represent an intriguing mathematical geometry, demonstrating relevance across scales ranging from atomic to cosmological dimensions. In the realm of nanophotonics, ring-shaped structures represent a key architype, particularly in photonic circuits,[1,2,3,4] quantum optics,[5,6] non-linear optics (NLO),[7,8,9] optical sensors,[10] and neuromorphic computing.[11] The prominence of microscale rings in photonic applications stems from their ability to function as resonators − structures that confine light and facilitate its circulation in clockwise and counterclockwise directions, thereby generating highly selective, narrow-bandwidth resonant optical modes.[12,13,14]

Nonetheless, the fabrication of micro ring resonators (MRRs) utilising organic crystalline materials poses a substantial challenge due to their inherently unfavourable crystal growth direction. To overcome these limitations, a surface-tension-assisted crystal growth technique has been developed to fabricate ring-shaped organic crystals.[15,16] Other reported approaches include molecular self-assembly,[12] mechanophotonics,[1,17] polydimethylsiloxane (PDMS) ring-hole stamp confinement method,[18] focused ion beam etching of organic crystal,[19] and polymer-based printing[4]. Despite these technological advances in creating ring-shaped materials, exploring their various mechanical motions is of particular interest, especially for applications in soft robotics, reconfigurable photonic circuits, and actuators. External stimuli-responsive molecular crystals garnered significant scientific attention due to the ability to exhibit diverse dynamic behaviours encompassing bending,[17,20] lifting,[17,21] twisting,[22,23] rolling,[24] curling,[25] crawling,[26] and jumping,[27] which have been extensively investigated. Nevertheless, achieving precise micromechanical control over organic MRRs and its NLO emissions remains underexplored. Development of next-generation organic NLO-MRRs that can translate mechanical stimuli into controlled microscopic motions represents a critical and timely research pursuit with substantial potential to advance diverse fields.

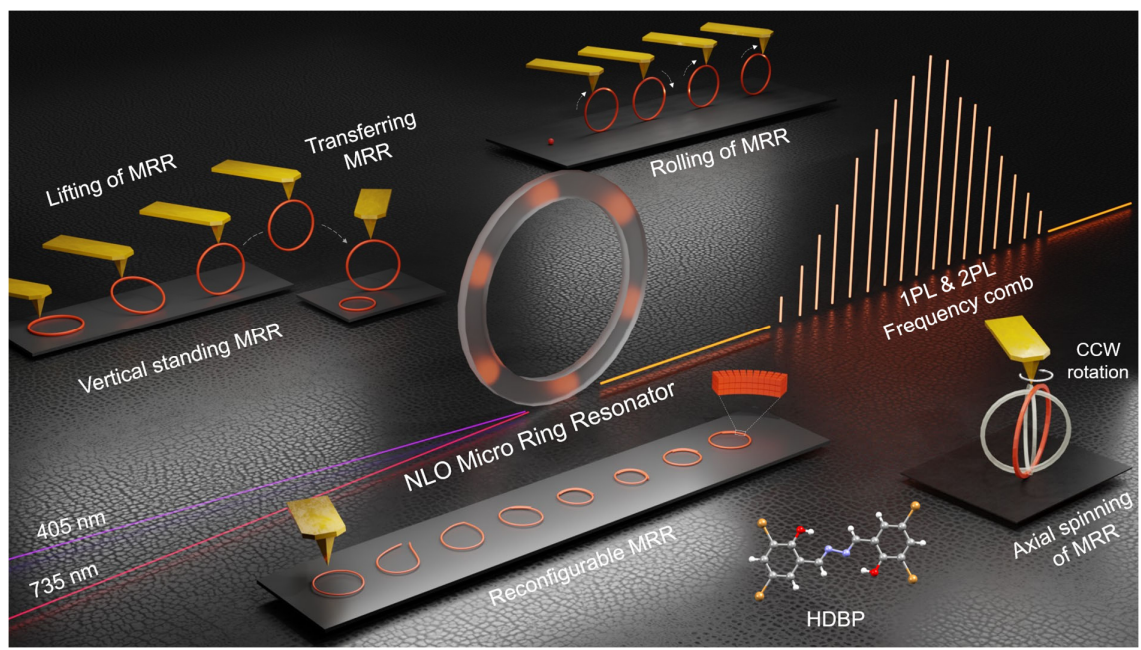

**Fig. 1 | Multifunctional and dynamic mechanical motions of MRR derived from HDBP.** Schematic illustration of the mechanically-driven multifunctional motions. (inset: molecular structure of HDBP.)

In this report, we present the fabrication of an organic MRR composed of 2,2'-((1*E*,1'*E*)-hydrazine-1,2-diylidenebis(methaneylylidene))bis(3,5-dibromophenol) (HDBP), through a surface-tension-assisted self-assembly technique. The MRR displays an orange two-photon luminescence (2PL) demonstrating its anisotropic NLO characteristics. Notably, the MRR displayed well-resolved optical whispering-gallery modes (WGMs) in both 1PL and 2PL regimes. We used mechanophotonics technique, an integrated setup that combines confocal microscopy to visualise microcrystals and an atomic force microscopy (AFM) cantilever tip to manipulate them in surface-bound and free-standing configurations. The pseudo-plasticity of a mechanically cut opened MRRs enables precise mechanical reconfiguration across various sizes and strain states, facilitating tailored optical paths and systematic optical resonance modulation. The MRRs exhibit excellent mechanical transferability, as they can be neatly detached from their original substrates and subsequently relocated onto new ones. Furthermore, these MRRs can be precisely positioned to stand vertically at various angles. Remarkably, these standing MRRs can be rolled like a wheel with translational precision and can also be axially rotated through a full 360°, all accomplished through a controlled guided movement of the AFM tip force. This extensive range of mechanical motions highlights the structural robustness and resilience of the MRRs derived from HDBP. This work bridges the gap between static optical components and dynamic, responsive systems that can reconfigure in real-time in response to mechanical stimuli, marking a significant leap in the development of mechanically controllable organic photonic devices (Fig. 1).

**Fabrication and optical performance of MRR**

The HDBP molecule was prepared via a modified synthetic route[28] using Schiff base condensation between 3,5-dibromosalicylaldehyde and hydrazine monohydrate (supplementary Figs. 1 and 2 and supplementary note. 1). The compound exhibits broad optical absorption up to around 470 nm and orange photoluminescence (PL) emission spanning from 540 nm to 770 nm with a maximum at approximately 625 nm (supplementary Fig. 3). The acicular HDBP crystals exhibit mechanical flexibility due to favourable weak intermolecular interactions including hydrogen bonding, halogen bonding, and π⋯π interactions. Notably, HDBP microcrystals retain deformed shapes under extreme strain, demonstrating *pseudo-plastic* behaviour, where strong substrate-crystal surface adhesion forces overcome crystal shape-restoring forces[28].

To prepare geometrically precise MRR using HDBP, we employed a surface-tension-assisted self-assembly technique[15](Fig. 2a). The HDBP MRRs are prepared by rapidly injecting a 1mM dichloromethane ($CH_2Cl_2$) solution of HDBP into Millipore water (1:10 volume ratio) under sonication to form a $CH_2Cl_2$-in-$H_2O$ emulsion. A 50 µL aliquot of the prepared $CH_2Cl_2/H_2O$ emulsion was then drop-casted onto a borosilicate glass substrate. Initially, $CH_2Cl_2$ microdroplets containing HDBP form and begin to shrink due to evaporation. Once the optimum size is reached, the HDBP solution disperses within microdroplets. As evaporation proceeds, surface tension drives the organic solution to the droplet edges, where molecular aggregation initiates nucleation sites, which drive them to grow into rod-like crystal morphologies due to one-dimensional crystal growth. Multiple nucleation sites with varying dimensions result in size variability in crystals. The curved droplet interface acts as a template, along with HDBP's flexible crystal lattice, and guides the microcrystals' growth into ring-like shapes. Complete ring closure occurs under optimal concentration, while facet-matching and suboptimal conditions yield semicircular configurations. The size of the microdroplet determines the diameter of the ring. Angular or lattice mismatches at the junction cause the crystal rod to grow either linear or curved trajectories, potentially forming multi-looped rings before solvent evaporation halts the growth process (supplementary Figs. 4 and 5 and supplementary videos 1 and 2).

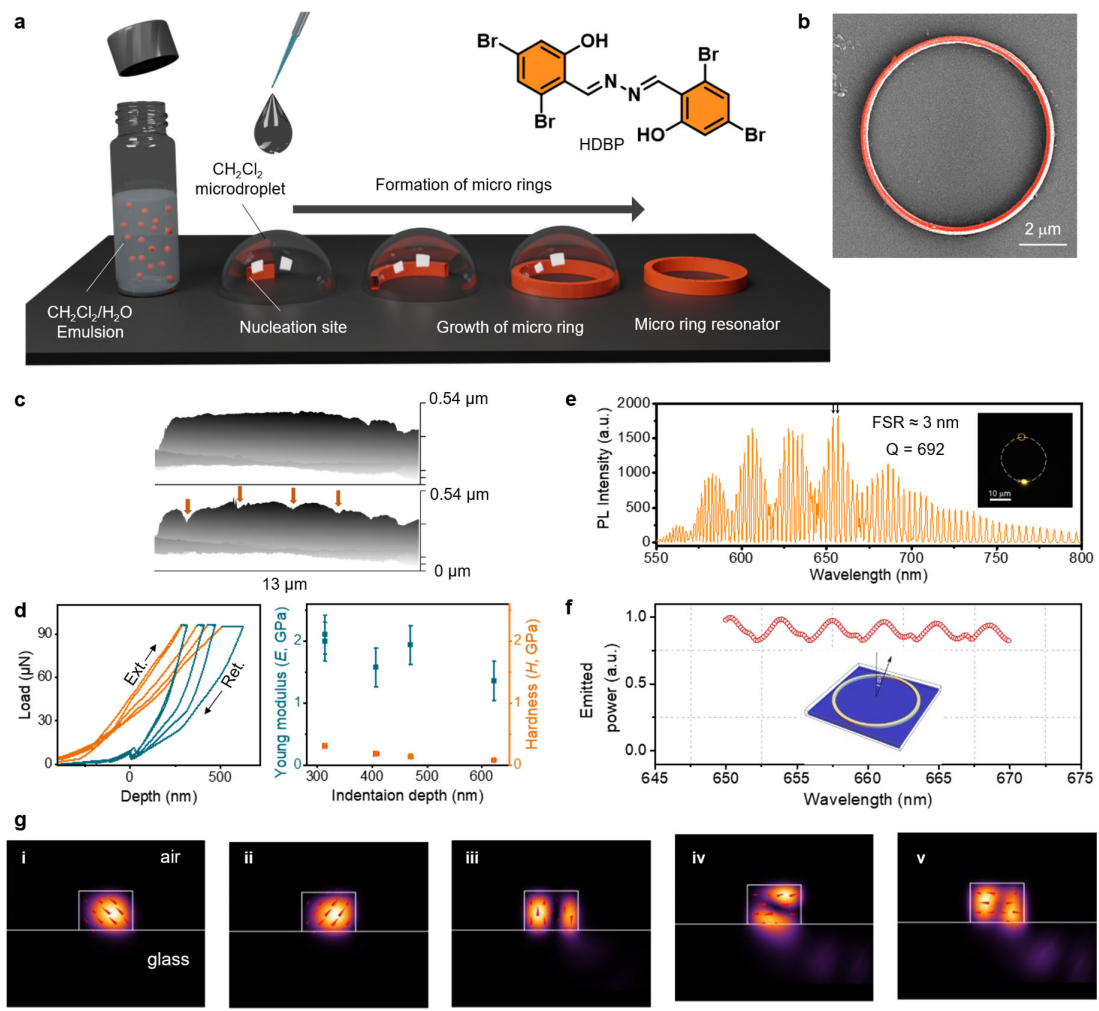

**Fig. 2 | Preparation, characterisation, and optical properties of organic MRRs. a** Schematic illustration of the fabrication methodology for MRRs. Molecular structure of HDBP is presented at the top right corner. **b** FESEM image of MRR. **c** AFM topography of the (001) plane recorded before and after nanoindentation (side-view) **d** Load (P) vs depth (d) curves obtained during nanoindentation performed on the (001) plane and corresponding plot of Young's modulus (*E*) and hardness (*H*) values of the MRR, respectively. **e** Background subtracted PL spectra of MRR1 (inset: PL image of excitation position. Open dot indicates the PL collection point. The white dashed circle denotes the light circulation pathway). **f** Spectrum of light power emitted from the cavity in the direction oriented at an angle of θ =100° (inset: Schematic representation of MRR1 on the glass substrate for FEM numerical calculation). **g** Spatial distribution of the squared amplitude of the optical electric field ($|E|^2$) for the eigenmodes labelled in plot supplementary Fig.10d. Red arrows indicate the direction of the electric field vector inside the cavity.

The field emission scanning electron microscopy (FESEM) image of MRR confirmed the perfect geometric precision of the formed microring structure (Fig. 2b). The AFM analysis of MRR1 (diameter $D_1 \approx 21.3$ µm) revealed height variation between 550-1000 nm, attributed to different size of nucleation sites which grown into a single ring (supplementary Figs. 6 and 7). The mechanical properties of MRR were characterized via AFM-based nanoindentation on major crystal planes. Load (µN) versus displacement (nm) curves were acquired for the (010) and (001) planes using a cone-shaped diamond tip (Fig. 2c, d and supplementary Fig. 8a-d). Minimal residual depth upon unloading demonstrated significant elastic recovery under mechanical stress. The mechanical properties calculated using the Oliver-Pharr method yielded Young's moduli of 1.796±0.312 GPa and

4.73±0.413 GPa for (010) and (001) planes, respectively, with corresponding hardness values of 0.204±0.030 GPa and 0.217±0.013 GPa (Fig. 2d and supplementary Fig.8e). The fluorescence lifetime imaging microscopic (FLIM) measurements of HDBP rings showed an average Fluorescence lifetime of 0.47 ns (supplementary Fig. 8). The PL image of the MRR1 displayed an intense orange emission under ultraviolet (UV) irradiation (supplementary Fig. 9). A confocal microscope in the transmission mode geometry was used to optically excite the MRR1 edge using a precisely focused (60x objective) continuous wave (CW) 405 nm diode laser, as the excitation wavelength lie within HDBP's absorption range. The resultant orange PL (540-800 nm) was recorded (20x objective) at the opposite edge of the MRR1 (inset of Fig. 2e). The multiple recirculation of PL within the crystal led to interference patterns with sharp WGM resonances resembling a frequency comb (Fig. 2e).

The free spectral range (FSR), defined as the wavelength difference between the consecutive resonant peaks (FSR = $\lambda_m$ - $\lambda_{m+1}$), was estimated to be about 3 nm for MRR1 (Fig. 2e). The FSR values vary with the ring diameter according to the relation, $m\lambda = \pi D n_{eff}$, where $\lambda$ is the wavelength, m the angular momentum mode number, D the diameter, and $n_{eff}$ the effective refractive index. Analysis of the WGM spectra of MRRs with diameters ranging from 4 µm to 35 µm showed a linear 1/D vs FSR trend, confirming that FSR increases as the MRR diameter decreases (supplementary Fig. 10a-j). The Q-factors, calculated as Q = $\lambda$/FWHM (FWHM = full width at the half maximum), reached ≈1000 for D≈35 µm, and were plotted against D to establish the relationship between the cavity size and optical confinement (supplementary Fig. 10k).

Finite element method (FEM) numerical calculations were performed for MRR1 (Fig. 2e-g). The geometric MRR parameters of width (W ≈ 809 nm), height (H ≈ 590 nm), refractive index of $n_r$ = 1.7+0.0013i, and $n_s$ (for glass substrate) = 1.51 were used for the calculations (inset of Fig. 3f). Optical losses were introduced with a nonzero value of the imaginary part of $n_r$ to match the experimental Q (≈700) (Fig. 2e). The calculated resonant wavelengths and Q-factors for different azimuth mode numbers (m) were determined using the FEM eigenmode calculation and are shown in supplementary Fig. 10d. The highest Q-factors for a given m-index correspond to modes with a single electric field antinode inside the cavity (points i and ii in supplementary Fig. 10d). The electric field distributions for m=160 labelled by roman numbers in supplementary Fig. 10d are shown in Fig. 2g. As can be seen, increasing the number of antinodes significantly reduces the Q-factors (see points iii, iv, and v in Fig. 2g). The low-order modes typically exhibit Q≈700, close to the experimental value. The calculated spectrum near λ=650 nm shows FSR ≈ 3.6 nm (supplementary Fig. 10e). The simulations of the spectrum of MRR1 were carried out by calculating the power emitted by a dipole-like light source situated inside the ring cavity. The far-field indicatrix of the scattered light (supplementary Fig. 11) for two wavelengths, corresponding to WGM resonance (red curve) and non-resonance (black curve) spectral ranges, shows enhanced emission at glide angles (θ ≈ 90°–110°) were used for further spectral calculations. At θ=100°, the spectrum showed a WGM-induced periodic modulation with an FSR of 3.6 nm, matching the eigenmode calculation (Fig. 2f). The modulation depth (about 18%) is limited by the finite Q-factor. In contrast, a lossless ring cavity ($n_r$ = 1.7) showed dramatically improved spectral resonances.

**Non-linear properties of MRR**

To investigate NLO effects and compare them with linear optical responses, MRR2 ($D_2$ ≈ 17.5 µm) was selected (Fig. 3a inset). For improved WGMs detection and minimised non-resonant PL, emission was recorded from the opposite side of the excitation spot (red stripe in Fig. 3a inset). The 1PL spectrum exhibited distinct WGMs with an FSR of ≈3.1 nm near 600 nm wavelength region after background subtraction (Fig. 3a). For 2PL, a 735 nm (200 fs, 80 MHz) laser was strategically chosen to be outside the linear absorption of HDBP while remaining within its two-photon absorption range. The pump beam was focused with a 100× objective (NA = 0.7) to a focal spot of approximately 0.7 μm in diameter, delivering a mean intensity of 15 kW/cm² that is well below MRR's damage threshold. As in 1PL, the 2PL emission was collected from the opposite side of the pumped part of the MRR to reduce non-resonant background. The identical optical setups enabled direct comparison of 1PL and 2PL spectral profiles.

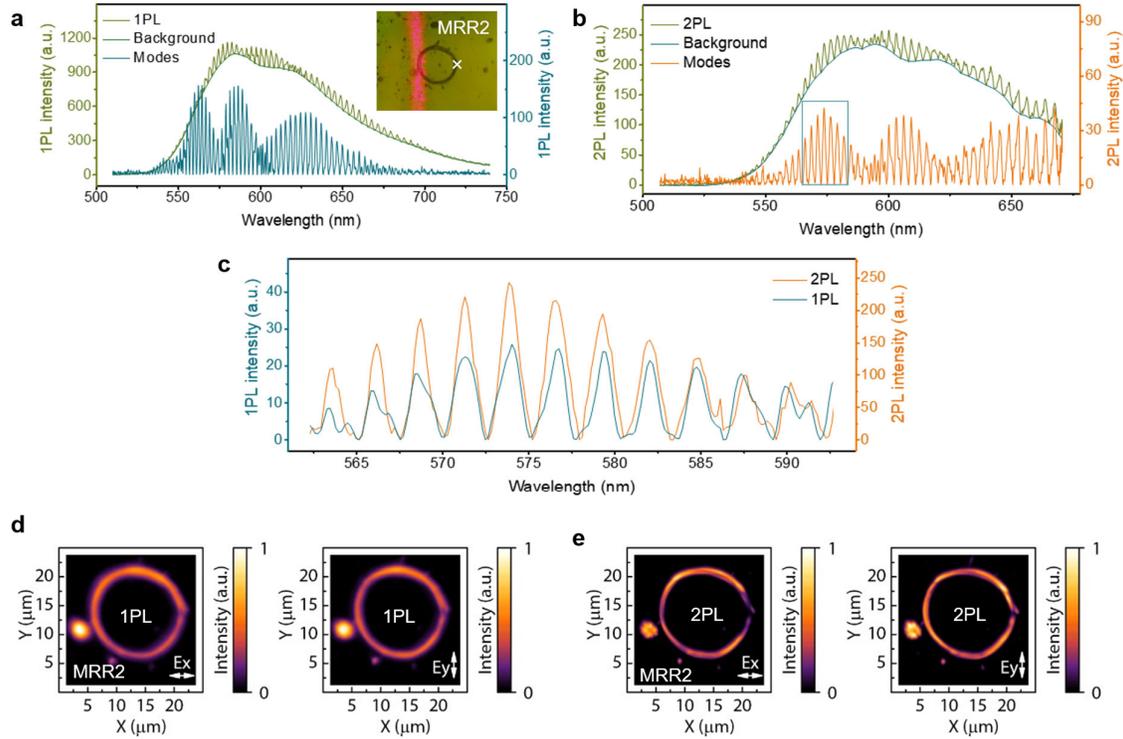

**Fig. 3 | Linear and non-linear optical properties of MRR. a** 1PL spectrum of MRR2 (inset: optical image of the MRR2 showing excitation (white cross mark) and collection position (red region)), **b** 2PL spectrum of MRR2. **c** Comparison of the 1PL and 2PL spectra of MRR2. **d, e** spatial maps of 1PL and 2PL of MRR2, respectively, measured for different pump polarizations.

The 2PL spectrum clearly exhibited WGM characteristics (Fig. 3b), demonstrating that MRRs act as effective NLO resonators. As expected, identical mode families appear in both 1PL and 2PL spectra (Fig. 3c), with excellent alignment of resonance peaks, indicating that the same WGM governs both processes. Spatial power mapping across MRR2 under linearly polarized excitation (white arrows) showed that 1PL intensity is polarization-independent (Fig. 3d), whereas 2PL exhibited subtle polarization dependence. Specifically, the brightest regions in the 2PL images shifted spatially depending on vertical or horizontal laser polarization (Fig. 3e). The second-order power dependence of 2PL on excitation intensity confirmed the NLO origin of the 2PL signal, displaying a reversible near-quadratic scaling after correcting for minor photobleaching (supplementary Fig. 12). Additionally, polarization-resolved measurements under 405 nm CW excitation revealed distinct transverse magnetic (TM) and transverse electric (TE) mode pairs with angular dependence on the polarizer azimuthal position. At 0°, TE modes dominated; at 90°, TM modes were predominant (supplementary Fig. 13).

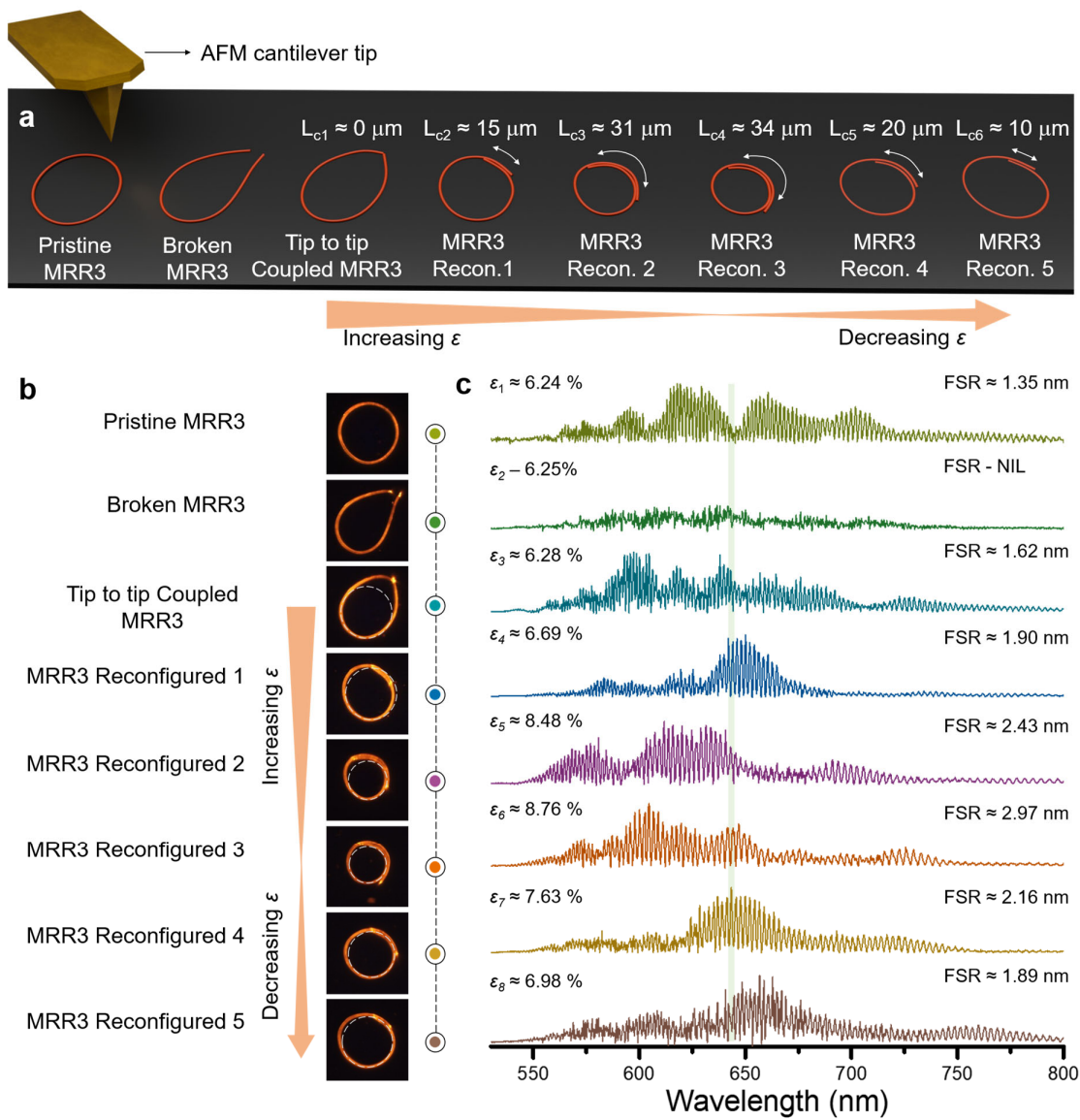

**Fig. 4 | Mechanically and optically Reconfigurable MRR. a** Graphical illustration demonstrating mechanical manipulation of reconfigurable MRR3. $L_c$ - length of coupling regions. **b** PL images of the reconfigurable MRR3 in each reconfigured state. **c** Corresponding stacked PL spectra. Insets: Left – corresponding strain ($\varepsilon$) values; Right – corresponding FSR values.

**Mechanically reconfigurable MRR**

Organic MRRs mechanically deposited on a substrate inherently possess a high strain value due to their curved geometry. Using mechanophotonics technique, their diameter can be dynamically reconfigured, thereby modulating strain ($\varepsilon$) and thus the FSR (Fig. 4a). We selected a pristine MRR3 ($D_3 \approx 33$ μm ; $\varepsilon_1 \approx 6.24\%$) that showed FSR of 1.35 nm. (supplementary Fig.4a, b) To validate the optical modes originating from the MRR3, the ring was deliberately cut open with an AFM cantilever. The PL spectrum of the broken ring showed no optical modes, confirming WGM formation requires a closed-loop ring geometry (Fig. 4b, and supplementary note 5, and supplementary Fig. 14). To verify further, the broken ring tips were precisely rejoined in a tip-to-tip configuration, restoring light circulation within the loop. The reconnected ring produced optical modes with a modified FSR value of 1.62 nm, attributed to a slight increase in strain ($\varepsilon_3 \approx 6.28\%$). Here, the coupling region length ($L_c$) was about 0 μm, enabling light re-entry into the loop via evanescent coupling between the crystal tips. To induce higher strains, bending mechanophotonic operation was applied using an AFM cantilever tip (supplementary note. 5). The MRR3 tips were repositioned to form overlapping regions, resulting in a non-circular geometry, with strain scaling with $L_c$. The MRR3 underwent three sequential reconfigurations: (1) $L_{c2} \approx 15$ μm, $\varepsilon_4 \approx 6.69\%$, FSR = 1.90 nm; (2) $L_{c3} \approx 31$ μm, $\varepsilon_5 \approx 8.48\%$, FSR = 2.43 nm; and (3) $L_{c4} \approx 34$ μm, $\varepsilon_6 \approx 8.76\%$, FSR = 2.97 nm. The FSR increased consistently with strain, confirming strain-tunable optical response from MRR3. To reverse the strain, further reconfiguration was performed to increase the size, leading to decreased FSR values: (4) $L_{c5} \approx 20$ μm, $\varepsilon_7 \approx 7.63\%$, FSR = 2.16 nm; and (5) $L_{c6} \approx 10$ μm, $\varepsilon_8 \approx 6.98\%$, FSR = 1.89 nm. The close match between the FSR of reconfiguration 1 and 5 confirms reversibility of ring deformation, highlighting its structural resilience and suitability for dynamic photonic device applications (Fig. 4b, c, and supplementary Figs. 14-17 and supplementary videos 3 and 4).

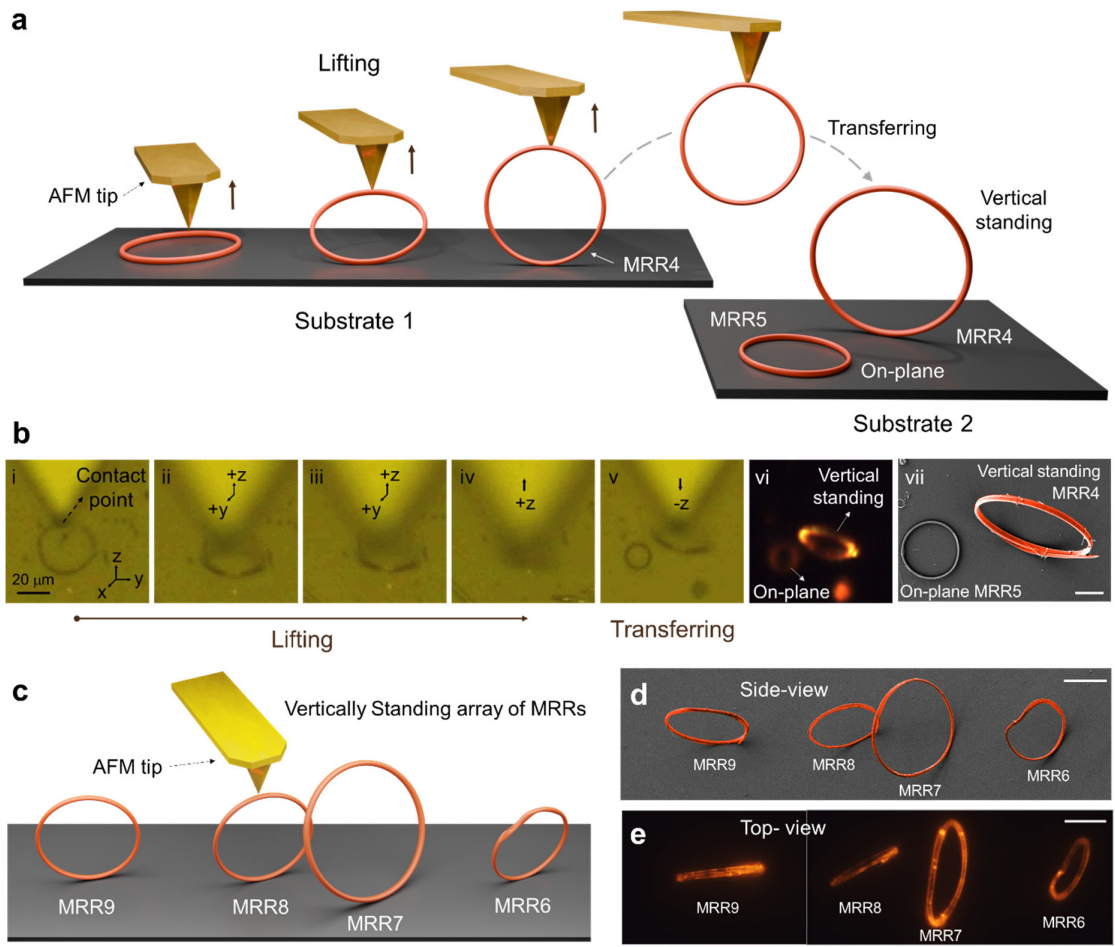

**Fig. 5 | Vertically standing MRRs. a** Graphical illustration of the MRR's vertical placement process and **b i-vi** Sequential confocal images of the lifting, transferring, and vertical standing mechanophotonic operations of MRR4. **b vii** Colour-coded FESEM micrograph of vertically standing MRR4 compared with on-plane MRR5. **c** Graphical 3D representation of four vertically standing MRRs **d** Colour-coded FESEM micrograph and **e** Merged PL image of vertically standing array of MRR6-MRR9. The scale bars are 10 μm.

**Vertical standing of MRR**

Next-generation 3D photonic circuits demand components operatable both on substrate and in air. To demonstrate this, an MRR4 ($D_4 \approx 33.8$ μm) was lifted using a modified AFM-based technique tailored to its circular geometry and strong substrate adhesion. The AFM tip was meticulously placed on the top side of the MRR4, making stable contact with the side face.

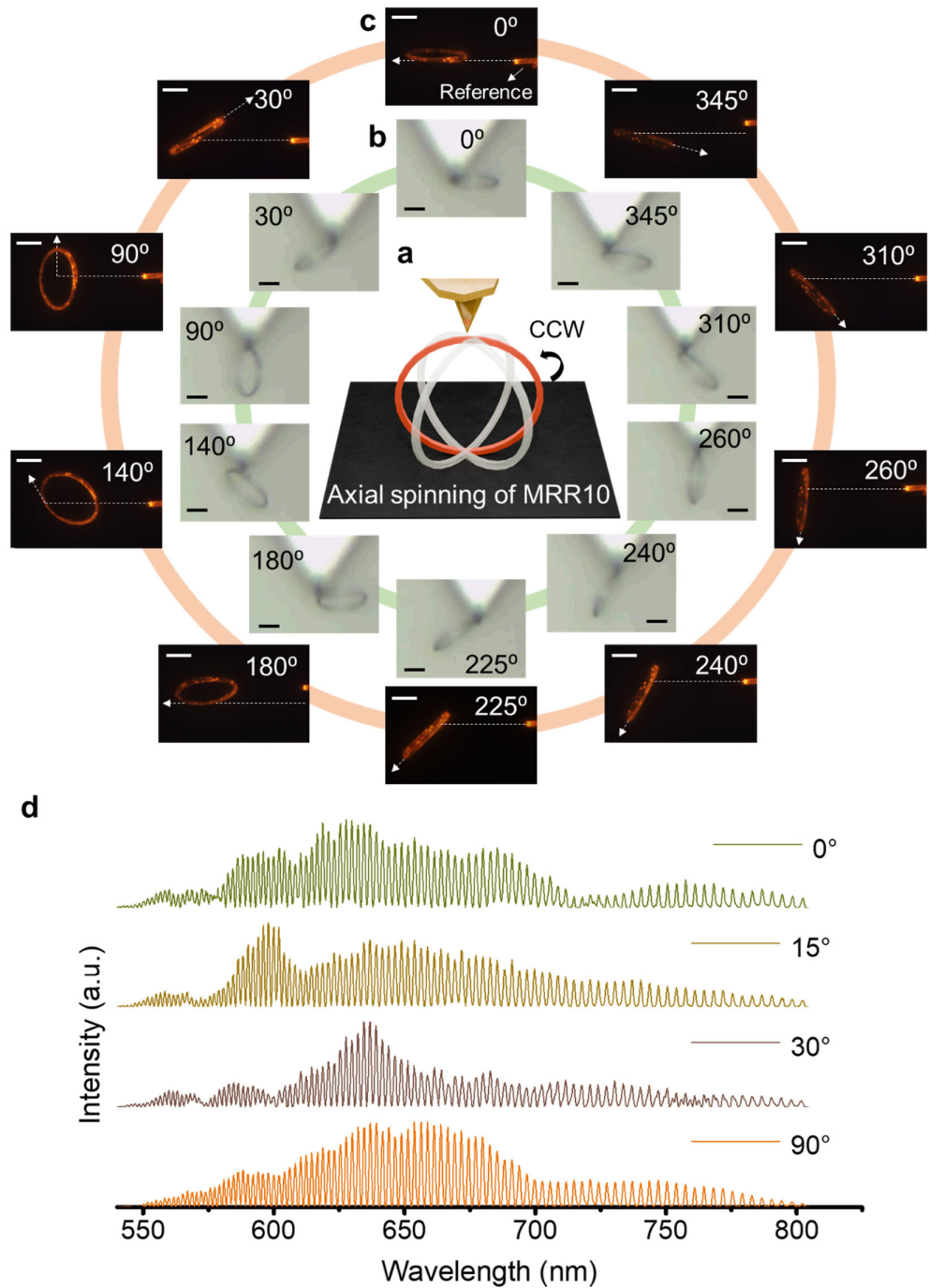

**Fig. 6 | Axial spinning of MRR. a** Graphical illustration demonstrating axial spinning of MRR10. **b** Confocal microscopy images and **c** PL images of the mechanical manipulation of the axial spinning of MRR10 at each angular orientation. The scale bars are 10 μm.

Afterwards, the tip was displaced forward (+y direction) to induce slight deformation, then upward (+z) to detach MRR4 from the substrate without damage and preserving its geometry (Fig. 5b). The lifted MRR4 was transferred and made to stand near MRR5, which lay on substrate plane, by precisely guiding the AFM tip downward (-z). The FESEM confirmed the vertically standing and the on-plane configurations (Fig. 5a, b and supplementary Fig. 18). To repeat this approach, four more MRRs (MRR6: $D_6 \approx 13.6$ μm; MRR7: $D_7 \approx 22.0$ μm; MRR8: $D_8 \approx 15.6$ μm; MRR9: $D_9 \approx 18.7$ μm) were similarly lifted and arranged as a vertically standing array on a different substrate. The PL and FESEM images confirmed the successful formation of a vertically standing MRR array. PL spectra collected from MRR6-MRR9 showed optical modes confirming that the photonic traits were fully retained in the vertical standing configurations (Fig. 5c-e and supplementary Figs. 18-21 and supplementary videos 5-7).

**Axial spinning of MRR**

Lab-on-chip actuator and soft-microrobotics require systems that translate mechanical stimuli into controlled motion. The MRRs provide such a platform, with axial spinning representing a previously unexplored operation, with the significant potential for advanced applications (Fig.6a). To demonstrate this, MRR10 ($D_{10} \approx 23.5$ μm) was lifted from the substrate and placed vertically using the previously described mechanophotonic procedure. A straight HDBP rod placed adjacent to MRR10 served as a 0° reference point for angular measurements. MRR10 was initially aligned in parallel orientation relative to the reference rod to establish the baseline configuration (Fig. 6b, c, panels 0°). For optical characterization, MRR10 was excited from below with a 405 nm laser, and spectra were collected from its top face. Applying the AFM cantilever tip's coordinated ±x and ±y movements, MRR10 was rotated at predefined angles, achieving full 360° spinning about its vertical axis. Optical spectra acquired at twelve distinct angles (0°-360°), provided comprehensive angular coverage throughout the complete rotational cycle (Fig. 6b, c). The spinning speed of the ring could be tuned for either rapid or gradual rotations, allowing for either rapid spinning motions or slow, controlled angular adjustments. The reproducibility and reliability of the axial spinning operation was confirmed by applying the same procedure to another MRR, validating axial spinning as a generalizable mechanophotonic approach for microring resonators (Fig. 6 and supplementary Figs. 22-25 and supplementary videos 8 and 9).

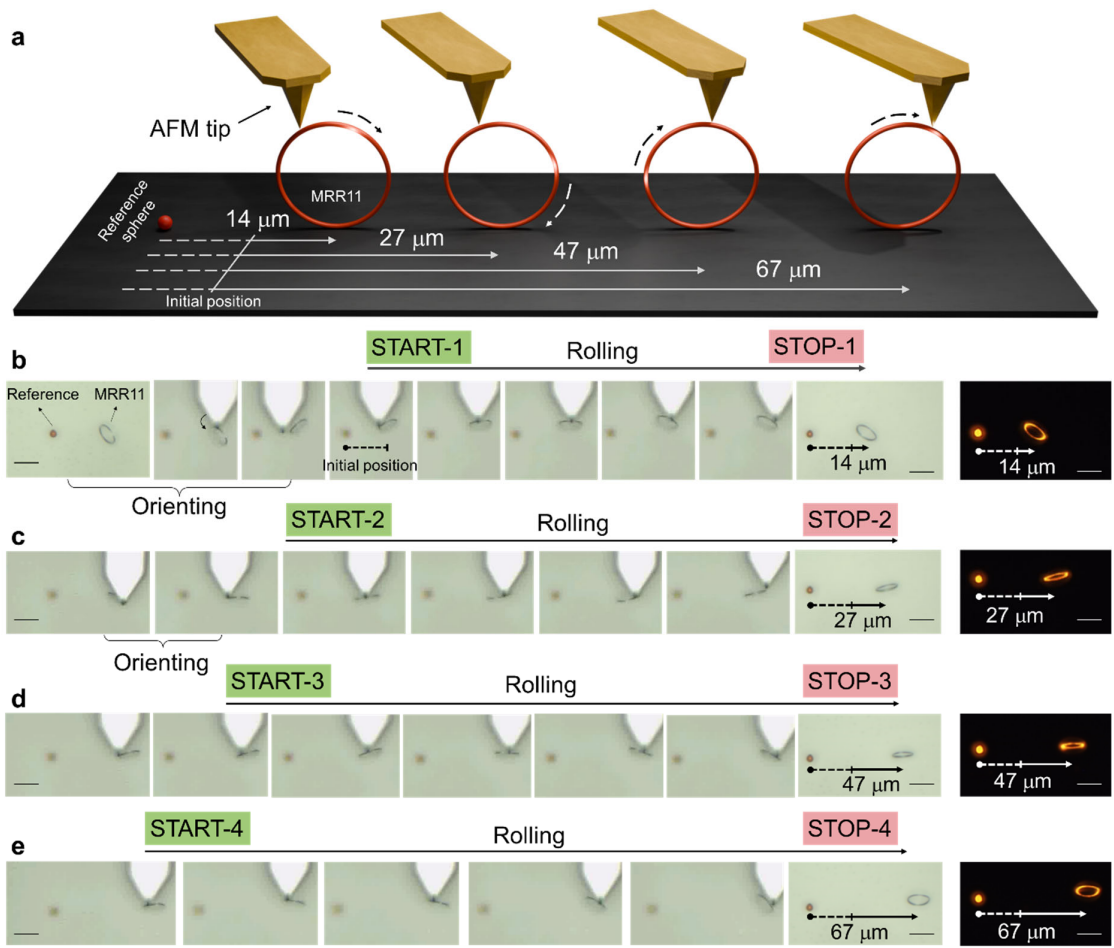

**Fig. 7 | Rolling locomotion of MRR11. a** Graphical representation demonstrating wheel-like rolling locomotion facilitated by AFM tip. Confocal image showing mechanical manipulation of MRR11 **b** from the initial point START-1 to the ending point STOP-1, and PL image of the final position after the first roll. **c** from the starting point START-2 to the ending point STOP-2, and PL image of the final position after the second roll. **d** from the starting point START-3 to ending point STOP-3 and PL image of the final position after the third roll. **E** from the starting point START-4 to ending point STOP-4 and PL image of the final position after the fourth roll. All the scale bars are 20 μm.

**Rolling locomotion of MRR**

To broaden the functional utility of MRRs, demonstrating their wheel-like rolling motion is essential (Fig. 7a). Therefore, MRR11 was lifted from its original substrate, vertically positioned onto another substrate. A HDBP-doped polymer sphere was placed ≈33.8 μm away as a reference marker for a rolling locomotion experiment. The initial position of MRR11 was marked, and the top face of the ring was optically focused. The MRR11 was aligned towards intended motion trajectory and the rolling locomotion was initiated by a +x movement of the AFM tip, gently contacting ring's top face. The initial rolling movement displaced MRR11 by 14 μm (Fig. 7b). Prior to each subsequent rolling operation, the ring was realigned to ensure proper rolling trajectory and prevent sliding or tumbling motions. The rolling locomotion was continued for three additional cycles, achieving progressive translational displacements of 27 μm, 47 μm, 67 μm, respectively (Fig. 7c-e). Optical spectra were collected at each stopping position throughout the rolling process, as in the axial spinning case. The spectra consistently exhibited sharp resonance modes at all measurement positions, providing definitive evidence that MRR11 retained its essential photonic properties despite multiple rolling operations. This establishes the remarkable mechanical robustness of the organic MRRs and their ability to retain optical functionality during dynamic rolling locomotion. For validation, another MRR was subjected to rolling locomotion. (Fig. 7 and supplementary Figs. 26-28 and supplementary videos 10 and 11).

**Conclusion**

This study demonstrates the distinctive dynamic characteristics of HDBP MRRs under various mechanical stimuli. Through comprehensive 2D and 3D mechanical manipulations, we systematically examined their influence on physical and optical properties. Our findings reveal MRR's exceptional adaptability for mechanically-driven soft machines and actuators. The preserved light-confining capabilities across all configurations underscore result reliability while amplifying photonic application potential. Additionally, the observed non-linear responses highlight their critical importance in contemporary photonic technologies, positioning HDBP MRRs as prime candidates for advanced nanophotonic applications.

**Methods**

Synthesis and fabrication of MRR

The compound was synthesized using a conventional Schiff base reaction using 3,5-dibromosalicylaldehyde and hydrazine monohydrate as starting materials. The micro ring geometry was achieved through an emulsifying process (DCM/Water). The resulting emulsion was drop-casted into a coverslip using a micropipette to obtain MRRs.

Optical studies

The optical performance was studied through a transmission mode setup of the Wi-Tec alpha laser confocal optical microscope (LCOM) equipped with a Peltier-cooled CCD detector. Groove frequency set to 300/mm and blaze wavelength to 750 nm. The system was configured with an integration time of 0.5 s, and the accumulation time was typically 10 s. Every single spectrum is obtained by ten averaged accumulations. The source of excitation was a diode 405 nm laser. For spectral measurements and image collection, a 20× and 150× objective was used. All measurements were performed under ambient conditions.

Mechanophotonics-mechanically driven motions of MRR

The micromanipulation experiments were performed using a confocal microscope with an AFM tip attached. The AFM cantilever was equipped manually with the 20x objective for all operations. Specifications of AFM cantilever are - Adama: NM-TC, with a force constant of 350 N/m, with a typical tip radius of 25±10 nm. The piezo-electric stage of the confocal microscope is the supporting platform of the coverslip containing MRRs, which was moved in the ±x and ±y directions after appropriately positioning the AFM cantilever tip near the crystal in the z-direction for the manipulation techniques. All operations were performed under ambient conditions.

# References

1. Zhang, C. et al. Organic printed photonics: From microring lasers to integrated circuits. *Sci. Adv.* **1**, (2015).

2. Chandrasekar, R. *Mechanophotonics for organic photonic integrated circuits*. (Institute of Physics Publishing, 2024).

3. Chandrasekar, R. Advanced All-Organic microphotonic components and integrated circuits. *Adv. Opt. Mater.* **11**, (2023).

4. Rohullah, M., Chosenyah, M., Kumar, A. V. & Chandrasekar, R. Cornu-Spiral-Like Organic Crystal Waveguide providing discriminatory optical pathway for smart organic photonic circuit. *Small* (2024).

5. Holzinger, R. et al. Harnessing quantum emitter rings for efficient energy transport and trapping. *Opt. Quantum* **2**, 57 (2024).

6. Silverstone, J. W. et al. Qubit entanglement between ring-resonator photon-pair sources on a silicon chip. *Nat. Commun.* **6**, (2015).

7. Gong, M. & Wu, H. Design of nonlinear optical ring resonators. *Conference on Lasers and Electro-Optics* JTh2A.28 (2019).

8. Black, J. A. et al. Optical-parametric oscillation in photonic-crystal ring resonators. *Optica* **9**, 1183 (2022).

9. Lin, J. et al. Flexible Organic Crystalline Fibers and Loops with Strong Second Harmonic Generation. *J. Am. Chem. Soc.* (2025).

10. Wang, J. et al. A silicon microring resonator for refractive index carbon dioxide gas sensing. *ACS Sens.* **10**, 4938–4944 (2025).

11. Srouji, L. E. et al. Photonic and optoelectronic neuromorphic computing. *APL Photonics* **7,** (2022).

12. Chandrasekhar, N. & Chandrasekar, R. Reversibly Shape-Shifting Organic Optical waveguides: formation of organic nanorings, nanotubes, and nanosheets. *Angew. Chem. Int. Ed* **51**, 3556–3561 (2012).

13. Bogaerts, W. et al. Silicon microring resonators. *Laser & Photonics Rev.* **6**, 47–73 (2011).

14. Wei, G., Wang, X. & Liao, L. Recent advances in organic Whispering-Gallery mode lasers. *Laser & Photonics Rev.* **14,** (2020).

15. Feng, X. et al. Spatially resolved organic Whispering-Gallery-Mode Hetero-Microrings for High-Security photonic barcodes. *Angew. Chem. Int. Ed.* **62,** (2023).

16. Lv, Y. et al. Controlled outcoupling of Whispering-Gallery-Mode lasers based on Self-Assembled organic Single-Crystalline microrings. *Nano Lett.* **19,** 1098–1103 (2019).

17. Chandrasekar, R. Mechanophotonics – a guide to integrating microcrystals toward monolithic and hybrid all-organic photonic circuits. *Chem. Commun.* **58,** 3415–3428 (2022).

18. Huang, H. et al. Wavelength-turnable organic microring laser arrays from thermally activated delayed fluorescent emitters. *ACS Photonics.* **6**, 3208–3214 (2019).

19. Pradeep, V. V., Chosenyah, M., Mamonov, E. & Chandrasekar, R. Crystal photonics foundry: geometrical shaping of molecular single crystals into next generation optical cavities. *Nanoscale* **15,** 12220–12226 (2023).

20. Yu, Q. et al. Photomechanical organic crystals as smart materials for advanced applications. *Chem. Eur. J.* **25,** 5611–5622 (2019).

21. Tahir, I. et al. Photomechanical crystals as Light-Activated Organic soft microrobots. *J. Am. Chem. Soc.* **146,** 30174–30182 (2024).

22. Kitagawa, D. et al. Control of photomechanical crystal twisting by illumination direction. *J. Am. Chem. Soc.* **140,** 4208–4212 (2018).

23. Tang, B., Yu, X., Ye, K. & Zhang, H. Manifold Mechanical Deformations of Organic Crystals with Optical Waveguiding and Polarization Rotation Functions. *Adv. Opt. Mater.* **10,** (2021).

24. Rohullah, M., Pradeep, V. V., Ravi, J., Kumar, A. V. & Chandrasekar, R. Micromechanically-Powered rolling locomotion of a Twisted-Crystal Optical-Waveguide cavity as a mobile light polarization rotor. *Angew. Chem. Int. Ed.* **61,** (2022).

25. Feng, H. et al. An Alloy Engineering Strategy toward Helical Microstructures of Achiral π-Conjugated Molecules for Circularly Polarized Luminescence. *PubMed* (2025).

26. Uchida, E., Azumi, R. & Norikane, Y. Light-induced crawling of crystals on a glass surface. *Nat. Commun.* **6,** (2015).

27. Takazawa, K. et al. Phase-transition-induced jumping, bending, and wriggling of single crystal nanofibers of coronene. *Sci. Rep.* **11,** (2021).

28. Kumar, A. V., Manoharan, D., Khapre, A., Ghosh, S. & Chandrasekar, R. An extremely Pseudo-Plastic, organic Crystal-Based Concentric-Ring-Resonator coupled optical waveguide. *Adv. Phys. Res.* (2024).

## Acknowledgements

R.C. thanks SERB (STR/2022/00011 and CRG/2023/003911), and DST/INT/RUS/RSF/P-71/2023(G) for funding. TM and VN acknowledge the financial support of the RSF grant 24-42-02009.

## Authors information

Authors and Affiliations

**Advanced Photonic Materials and Technology Laboratory, School of Chemistry, University of Hyderabad, Gachibowli, Hyderabad 500046, Telangana, India.**

Melchi Chosenyah, Swaraj Rajaram, Ankur Khapre, Rajadurai Chandrasekar

**Center for Nanotechnology, University of Hyderabad, Gachibowli, Hyderabad 500046, Telangana, India.**

Rajadurai Chandrasekar

**Quantum Electronics Division, Department of Physics, Lomonosov Moscow State University, Leninskie Gory 1, Moscow 119991, Russia.**

Vladimir Novikov, Anton Maydykovskiy, Tatiana Murzina

Contributions

M.C. and S.R. synthesised, characterized and performed optical studies on the MRR under the supervision of R.C. M.C. performed the mechanical manipulation studies on MRR under the supervision of R.C. V.N. carried out the

FEM numerical calculations under the supervision of T.M. A.M. performed non-linear optical studies. A.K. carried out the AFM-based nanoindentation studies under the supervision of R.C. The paper was written with contributions from all authors. All authors have approved the final version of the paper.

Corresponding author

Rajadurai Chandrasekar

**Ethics declarations**

Competing interests

The authors declare no competing interests.